\documentclass[pre,floatfix,twocolumn,showpacs]{revtex4}
\usepackage{amsmath,bm}
\usepackage{graphicx}

\renewcommand{\vec}[1]{\mbox{\boldmath $#1$}}

\def\beg{\begin{eqnarray}}
\def\ende{\end{eqnarray}}
\def\lsim{\lower.4ex\hbox{$\;\buildrel <\over{\scriptstyle\sim}\;$}}

\newcommand{\nab}{\mbox{\boldmath $\nabla$} {}}

\newcommand{\Eq}[1]{Eq.~(\ref{#1})}
\newcommand{\EQ}{\begin{equation}}
\newcommand{\EN}{\end{equation}}

\newcommand{\ff}{{\bm{f}}}
\newcommand{\uu}{{\bm{u}}}
\newcommand{\BB}{{\bm{B}}}
\newcommand{\JJ}{{\bm{J}}}
\newcommand{\UU}{{\bm{U}}}

\newcommand{\meanB}{\overline{B}}
\newcommand{\meanBB}{\overline{\vec{B}}}
\newcommand{\meanJJ}{\overline{\vec{J}}}
\newcommand{\aT}{{a}^{\rm T}}
\newcommand{\aaT}{{\bm{a}}^{\rm T}}
\newcommand{\bbT}{{\bm{b}}^{\rm T}}
\newcommand{\meanBT}{{\overline{B}}^{\rm T}}
\newcommand{\meanBBT}{{\overline{\vec{B}}}^{\rm T}}
\newcommand{\meanJT}{{\overline{J}}^{\rm T}}
\newcommand{\meanJJT}{{\overline{\vec{J}}}^{\rm T}}

\newcommand{\meanUU}{\overline{\vec{U}}}
\newcommand{\SSS}{{\sf S}}
\newcommand{\SSSS}{\mbox{\boldmath ${\sf S}$} {}}
\newcommand{\urms}{u_{\rm rms}}
\newcommand{\kf}{k_{\rm f}}
\newcommand{\cs}{c_{\rm s}}
\newcommand{\etaT}{\eta_{\rm T}}

\newcommand{\DD}{{\rm D}}
\newcommand{\Rm}{{\rm Rm}}
\newcommand{\DDD}{{\cal D}}
\newcommand{\xxx}{\hat{\bm{x}}}
\newcommand{\yyy}{\hat{\bm{y}}}
\begin{document}

\title{The alpha effect  in a turbulent liquid-metal   plane Couette flow}
\author{G. R\"udiger}
     \affiliation{Leibniz Institute for Astrophysics  Potsdam,
         An der Sternwarte 16, D-14482 Potsdam, Germany}
 \affiliation{Helmholtz-Zentrum Dresden-Rossendorf,
        P.O. Box 510119, D-01314 Dresden, Germany}
\email{gruediger@aip.de}
\author{A. Brandenburg}
\affiliation{Nordita, KTH Royal Institute of Technology and Stockholm University,
Roslagstullsbacken 23, SE-10691 Stockholm, Sweden}
\affiliation{Department of Astronomy, Stockholm University, SE-10691 Stockholm, Sweden}

\date{\today}

\begin{abstract}
We calculate the mean electromotive force in plane Couette flows of
a nonrotating conducting fluid under the influence of a large-scale
magnetic field for driven turbulence.
A vertical stratification of the turbulence intensity results in an
$\alpha$ effect owing to the presence of horizontal shear.
Here we discuss the possibility of an experimental determination of
the components of the $\alpha$ tensor using both quasilinear theory and
nonlinear numerical simulations.
For magnetic Prandtl numbers of the order of unity, we find that
in the high-conductivity limit the $\alpha$ effect in the direction of
the flow clearly exceeds the  component in spanwise direction.
In this limit, $\alpha$ runs linearly with the magnetic Reynolds number
$\rm Rm$ while in the low-conductivity limit it runs with the product
$\rm Rm\cdot Re$, where $\rm Re$ is the kinetic Reynolds number so that
for given $\rm Rm$ the $\alpha$ effect grows with decreasing
magnetic Prandtl number.

For the small magnetic Prandtl numbers of liquid metals,
a common value for the horizontal elements of the $\alpha$ tensor appears,
which makes it unimportant whether the $\alpha$ effect is measured in
the spanwise or streamwise directions.
The resulting effect should lead to an observable voltage in both directions
of about 0.5\,mV for magnetic fields of 1\,kgauss and velocity fluctuations
of about 1\,m/s in a channel of 50\,cm height (independent of its width).
\end{abstract}


\maketitle
\section{Introduction}
Mean-field electrodynamics of turbulent conducting fluids provides the
commonly accepted approach to explaining the existence of magnetic fields
of cosmic bodies.
The excitation of magnetic fields results from the interplay of two
elementary processes, diffusive and non-diffusive ones.
It is known that turbulent motions reduce large-scale electric
currents by inducing an electromotive force (EMF) opposite to the
direction of the current. One can write 
\begin{equation}
\overline{\vec{u} \times \vec{b}} = - \mu_0 \eta_{\rm T} \meanJJ,
\label{EMF}
\end{equation}
where $\vec{\cal E}=\overline{\vec{u} \times \vec{b}}$ is the EMF with
$\vec{u}=\UU-\meanUU$ and $\vec{b}=\BB-\meanBB$ being the
fluctuating contributions to velocity $\UU$ and magnetic
field $\BB$, respectively, overbars denote averaging
(to be specified later), $\mu_0$ is the vacuum permeability,
$\JJ={\rm rot}\BB/\mu_0$ is the current density,
and $\eta_{\rm T}$ is the turbulent magnetic diffusivity.
In stellar convection zones, $\eta_{\rm T}$ exceeds the molecular
(microphysical) value $\eta$ of the magnetic diffusivity by many orders of magnitudes.

In this paper the EMF is derived for a turbulent fluid in the presence of a
mean shear flow $\meanUU$ and a {\em uniform} background field $\meanBB$.
For a turbulent dynamo, the enhanced dissipation must be overcome
by an induction process that does not run with the electric current.
One also knows that under the influence of global rotation and a  uniform magnetic field, anisotropic turbulence produces an EMF parallel to the field \cite{KR80}, i.e.\
\begin{equation}
\overline{\vec{u} \times \vec{b}} = \alpha \meanBB - \dots .
\label{EMF1}
\end{equation}
Here, $\alpha$ is a pseudoscalar formed by the rotation vector $\vec{\Omega}$
and the anisotropy direction $\vec{g}$
with $\alpha \propto \vec{g} \cdot\vec{\Omega}$.
Often, the anisotropy direction from the gradient of density stratification
of the fluid is used, but it is also possible that an intensity gradient
close to rigid boundaries forms the preferred direction.

The resulting dynamo equation in rotating and stratified plasma is \cite{KR80}
\begin{equation}
\frac{\partial\vec{\bar B}}{\partial t}= {\rm rot}\bigg(\alpha \vec{\bar B} - (\eta+\eta_{\rm T}) {\rm rot} \vec{\bar B}\bigg),
\label{Bt}
\end{equation}
which has non-decaying solutions if $\alpha$ exceeds a critical value
(\lq $\alpha^2$-dynamo'); see Ref.~\cite{B05} for a review.

In many papers the presented concept of a turbulent dynamo has been
applied to planets, stars, accretion disks, galaxies and galaxy clusters;
see references in \cite{RH04}.
Only very few papers, however, deal with an experimental confirmation of the validity of relations (\ref{EMF}) and (\ref{EMF1}) in the laboratory.
This is surprising given the astrophysical importance of Eq.~(\ref{Bt})
as a direct consequence of Eqs.~(\ref{EMF}) and (\ref{EMF1}), which
characterizes the basic ingredients of electrodynamics
in rotating turbulent fluid conductors.
Generally, the validity of Eq.~(\ref{EMF}) is not seriously doubted.
However, the existing laboratory experiments report an increase of the
effective magnetic diffusivity by only a few percent \cite{St12}.
This is because the molecular magnetic diffusivity is rather large
and the turbulence not strong enough.
This is an unfortunate situation as the eddy-concept of the effective dissipation in turbulent media governs much of cosmic physics from climate research,
geophysics, to the theory of star formation and quasars.

An even more dramatic situation holds with respect to the $\alpha$ effect.
There are one or two experiments on the basis of the idea that
the $\alpha$ effect is essentially a measure of the swirl of the flow.
It has been demonstrated that a fluid with imposed helicity
(imposed by rigid, swirling channels), produces an EMF in the direction
of an imposed field; see Refs.~\cite{St67,St06,F08}.
It is not yet shown, however, that a rotating fluid with helicity that
is {\em not} imposed (but results from the global rotation of the fluid)
leads to an observable $\alpha$ effect.
In natural cosmic bodies, helicity is usually due to the interaction
of rotating turbulence with density stratification.
The aim of the present paper is to suggest such a more rigorous
$\alpha$ effect experiment.
As we shall demonstrate, the difficulties in such an experiment
make it understandable that this has not yet been possible without
the use of a prescribed helicity.

It is easy to see the general difficulty of performing
$\alpha$ effect experiments.
Using Eq.~(\ref{EMF1}), the potential difference between the endplates
of the container in the direction of the mean magnetic field is
\begin{equation}
\Delta \varPhi= \alpha \meanB H,
\label{delU}
\end{equation}
where $H$ is the distance between the endplates.
Hence, for $H\simeq 100$ cm and $B\simeq 1000$ gauss (say)
the potential difference is $\Delta {\varPhi}=1$\,mV for
$|\alpha|=1\,$cm/s.
The maximum $\alpha$ value is of the order of $u_{\rm rms}$, hence $\Delta
{\varPhi} \lsim  u_{\rm rms}$ in mV. For $u_{\rm rms} \simeq 1$\,cm/s the
maximally induced potential difference is therefore 1\,mV.
A container of 5 cm radius rotating with 1 Hz  has a linear outer
velocity of more than 30 cm/s so that $u_{\rm rms} \simeq 1$ cm/s might
be considered as a conservative estimate.
We find as a necessary condition for any $\alpha$ experiment  that one
must be able  to measure  potential differences {\em smaller} than a
few mV.
The $\alpha$ experiment in Riga \cite{St67}
worked with $B\simeq 1$ kgauss and velocities of the order of m/s,
so that the $\Delta {\varPhi}$ exceeded 10 mV.
This experiment, however, used a prescribed helical geometry to mimic
the symmetry breaking between left- and right-handed helicities.

If the rotation is not uniform, the resulting shear
induces toroidal magnetic fields so that for sufficiently strong shear
the $\alpha$ effect can be rather small and still produce a dynamo
(\lq $\alpha\Omega$-dynamo').
It is well known that also turbulence in liquid metals subject to
a plane shear flow (without rotation!) is able to work as a dynamo if
the turbulence intensity is stratified in the  direction orthogonal to
the shear flow plane; see Ref.~\cite{RK06}. 
The basic rotation may thus not be the only flow, whose influence
enables the turbulence to generate global magnetic fields.

In the present paper, a plane Couette flow is considered to
analyze the characteristic issues of the corresponding $\alpha$ effect  and to design  a possible  experiment to measure its amplitude. 
\section{The mean electromotive force}
Consider  a plane shear flow with 
uniform vorticity in the vertical $z$ direction, i.e.,
\beg
{\bar U}_y=S x,
\label{1}
\ende
where $S$ is the shear rate.
The shear flow may exist in a turbulence field that does not
possess anisotropy other than that induced by the shear
(\ref{1}) itself. The one-point correlation tensor is
 \beg
 Q_{ij}=\overline{ u_i(\vec{x},t) u_j(\vec{x},t)}. 
\label{2}
 \ende
 The correlation tensor  may be constructed by a perturbation method.
The fluctuating velocity field is represented by a series expansion,
\beg
  {\vec u} = {\vec u}^{(0)} + {\vec u}^{(1)} + {\vec u}^{(2)} + ...\ ,
  \label{7}
\ende
where the upper index shows the order of the  contributions  in terms of the mean shear flow.

The zero-order term represents the \lq original' isotropic turbulence,
which is assumed as being not yet influenced by the shear.
We denote the Fourier transform of the correlation tensor by a hat
and define the spectral tensor for the original turbulence as
 \beg
 \hat Q_{ij}^{(0)}=\frac{E(k,\omega)}{16\pi k^2} \left(\delta_{ij}-\frac{k_i
 k_j}{k^2}\right),
 \label{8}
 \ende
where the positive-definite spectrum $E$ gives the intensity of isotropic
fluctuations with
\beg
\overline{ {\vec{u}^{(0)}}^2}=\int\limits_0^\infty\!\!\int\limits_0^\infty E(k,\omega) \
{\rm d}k \ {\rm d}\omega .
\label{9}
\ende
Here, $\vec{k}$ and $\omega$ are wavevector and frequency.
For analytical calculations, the one-parametric spectrum
\beg
E(k,\omega ) = \frac{2}{\pi}\frac{w}{\omega^2 + w^2}\hat E(k) ,
\label{18}
\ende
can be used, which yields a $\delta$ function, $E \propto \delta(\omega )$,
in the limit of $w\rightarrow 0$
and it leads to a white noise spectrum for large $w$. The correlation time of the turbulence is defined as $\tau_{\rm corr}=1/w$.
The extremely short correlation times of white noise automatically lead
to the high-conductivity limit for all fluid conductors with finite
magnetic diffusivity $\eta$.
On the other hand, the application of (\ref{18}) in form of a
$\delta$ function provides the result in the low-conductivity limit.

By definition, the magnetic diffusivity tensor relates the mean electromotive
force (\ref{EMF}) to gradients of the mean magnetic field via the relation
${\cal E}_i=\eta_{ijk}  \bar{B}_{j,k}$.
This tensor for originally isotropic turbulence,
influenced by a mean shear flow (\ref{1}), has been constructed
up to the first order in the shear \cite{RK06,RS06}.
In that work, it was also shown that the combination of shear and
the shear-induced parts of the magnetic diffusion tensor are not able
to operate as a dynamo.

On the other hand, it has been shown in Ref.~\cite{RK06} that shear,
in combination with {\em stratified} turbulence, provides helicity
that leads to an $\alpha$ effect in Eq.~(\ref{EMF1}).
Here, $\alpha$ must be a pseudotensor so that an $\epsilon$ tensor
has to appear in the coefficients for $\alpha$.
The construction of the EMF, ${\cal E}_i = \epsilon_{ijk}\overline{ u_j b_k}$,
is the only possibility for the $\epsilon$ tensor to appear.
The subscript of ${\cal E}_i$ is therefore always also a subscript
of the $\epsilon$ tensor.
As the $\epsilon$ tensor is of rank 3, an inhomogeneity of turbulence
with the stratification vector $\vec{g}=\nab\log u_{\rm rms}^2$ and
$u_{\rm rms}=\sqrt{\overline{u^2}}$, must also
be present for the $\alpha$ effect to exist.
If shear is included to first order, the general
structure of the $\alpha$ tensor is
 \beg
  \alpha_{ij} \!\!&=&\!\! \gamma \epsilon_{ijk}  g_k + \big(\alpha_1
  \epsilon_{ikl} \bar{U}_{j,k} +
 \alpha_2 \epsilon_{ikl}  \bar U_{k,j} \big) g_l +\nonumber\\ \!\!&+&\!\!
\alpha_3
 \epsilon_{ikl} g_j   \bar U_{l,k} + \alpha_4 \epsilon_{ikj} \bar U_{l,k}
 g_l + \alpha_5 \epsilon_{ijk} \bar U_{k,l} g_l .  
 \label{34} 
 \ende
If the stratification is along the vertical $z$ axis, it follows from
(\ref{34}) for the horizontal components of the $\alpha$ tensor  that
 \beg
  \alpha_{xx}&=& \alpha_2 g_z S= \alpha_x S,\nonumber\\ \alpha_{yy}&=&
  -\alpha_1 g_z S= \alpha_y S, \nonumber\\
\alpha_{xy}&=&-\alpha_{yx}   =\gamma g_z =\Gamma.
    \label{35}
 \ende
Turbulent pumping is characterized by $\alpha_{xy}$.
The anisotropy of the $\alpha$ tensor is described by the difference
between $\alpha_x$ and $\alpha_y$.
In the adopted geometry, the azimuthal component $\alpha_{yy}$
(the coefficient $\alpha_1$ defined below) plays the main role
in all cosmic applications, while in the proposed experiment with a
turbulent shear flow the coefficient $\alpha_2$ is probed,
and it produces the EMF {\em perpendicular} to the flow.
 
The coefficients in (\ref{35}) read
 \beg
  \gamma = \frac{1}{6}\int\limits_0^\infty\!\!\int\limits_0^\infty
  \frac{\eta k^2 E(k,\omega)}{\omega^2+\eta^2 k^4}
   {\rm d}k \ {\rm d}\omega
   \label{36}
 \ende
for the pumping term and
 \beg
  \alpha_n = \int\limits_0^\infty\!\!\int\limits_0^\infty
  A_n E\left( k,\omega \right) {\rm d}k \ {\rm d}\omega ,
  \label{37}
 \ende
for the $\alpha$ effect with
 \beg
  A_1 &=& \frac{4\nu\eta^3k^8 + 2\omega^2\eta \left(\nu + \eta\right)k^4}
  {15 \left(\omega^2+\nu^2k^4\right)\left(\omega^2 +
  \eta^2k^4\right)^2} +
  \nonumber \\
  &+& \frac{\eta^2k^4\left(\eta^2k^4 - 3\omega^2\right)}
  {15\left(\omega^2 + \eta^2k^4\right)^3} ,
  \nonumber \\
  A_2 &=& -\frac{\eta^2\nu^3\left(4\eta-5\nu\right)k^{12}}
  {60 \left(\omega^2+\nu^2k^4\right)^2\left(\omega^2 +
  \eta^2k^4\right)^2}-
  \nonumber\\
  &-&\frac{\omega^2\nu\left(28\eta^3 - 4\eta^2\nu + 12\eta\nu^2
  + 5\nu^3\right)k^8}
  {60 \left(\omega^2+\nu^2k^4\right)^2\left(\omega^2 +
  \eta^2k^4\right)^2} -
  \nonumber\\
  &-&\frac{\omega^4\eta\left(\eta + 36\nu\right)k^4 - 5\omega^6}
  {60 \left(\omega^2+\nu^2k^4\right)^2\left(\omega^2 +
  \eta^2k^4\right)^2}
  \label{38}
 \ende
for the kernels.
Here, $\nu$ is the kinematic viscosity.
In the following we define the magnetic Prandtl number as ${\rm Pm}=\nu/\eta$.
Only the terms occurring in (\ref{35}) have been given.
For small ${\rm Pm}$, one can easily estimate the coefficient $\alpha_1$.
For $\nu\ll \eta$, the expression for $A_1$ simplifies to
\beg
A_1&=& \frac{1}{15} \frac{1}{\left(\omega^2+\eta^2 k^4\right)^2} \Bigg\{\frac{4\nu\eta^3k^8}{\omega^2+\nu^2 k^4}+  \nonumber\\
&+&  \frac{\eta^2 k^4 \left(3\eta^2 k^4 - \omega^2\right)}{\omega^2+\eta^2 k^4}\Bigg\}  .
\label{39}
\ende
For $\nu \to 0$, the first expression on the RHS forms a $\delta$ function.
Hence,
\beg
\alpha_1=\frac{2\pi}{15 } \int\limits_0^\infty \!\!  \frac{E (k,0)}{\eta k^2} {\rm d}k + \dots
 \simeq \frac{2 \pi}{15} {\rm Rm}\ \ell_{\rm corr}^2 + \dots,
\label{41}
\ende
with the magnetic Reynolds number of the turbulence
\beg
{\rm Rm}= \frac{u_{\rm rms}^2 \tau_{\rm corr}}{\eta}.
\label{RmDef}
\ende
The missing terms in (\ref{41}), however, are  of the same order
as the given one, so that it can only be used for orientation. In (\ref{41}) we have used the estimate 
\beg
 \int\limits_0^\infty \!\!  \frac{E (k,0)}{k^2} {\rm d}k =   \tau_{\rm corr} u^2_{\rm rms} \ell_{\rm corr}^2,
\label{411}
\ende
which follows from (\ref{18}).

\begin{figure}[htb]
\vbox{ \includegraphics[width=8cm]{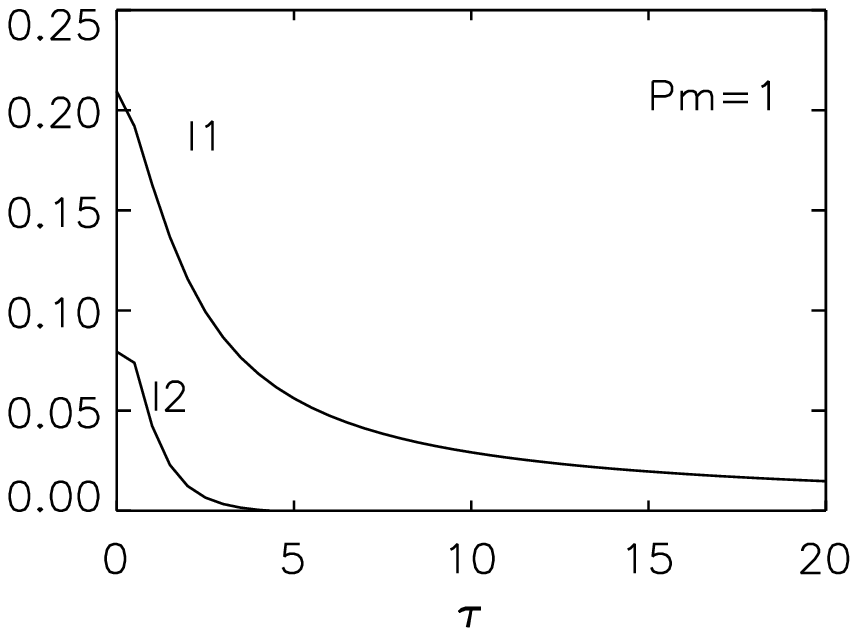}
\includegraphics[width=8cm]{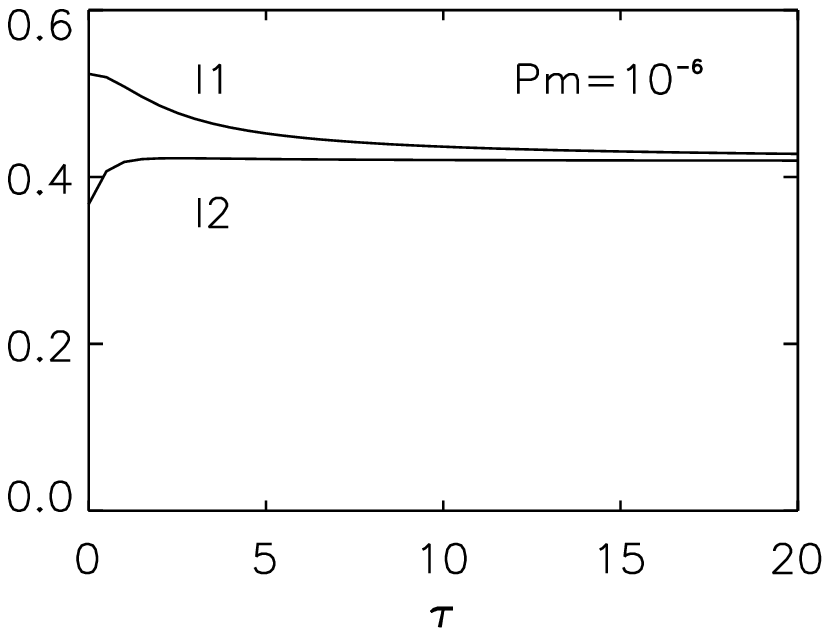}
}
\caption{The numerical values of the coefficients $I_1$ and $I_2$.
Top: $\rm Pm=1$, bottom:  $\rm Pm=10^{-6}$.
Note that in the high-conductivity limit ($\tau\to 0$, white noise spectrum) the coefficients
$I_1$ and $I_2$ for $\rm Pm\ll 1$ exceed the values for  $\rm Pm= 1$. For small $\rm Pm$ the  differences between  $I_1$ and $I_2$ and the influence of the diffusivity parameter $\tau$ almost vanish.  
}
\label{f1}
\end{figure}

As mentioned above, the white-noise approximation mimics the
high-conductivity limit, which holds for cosmic applications.
In this approach, the spectrum does not depend on the frequency $\omega$
up to a maximum value $\omega_{\rm max}$, above which the power
spectrum vanishes.
This corresponds to a turbulence model with very short correlation time,
i.e., $\tau_{\rm corr}\simeq 1/\omega_{\rm max}$.
One finds from (\ref{39}) after integration
\beg
\alpha_1=\frac{\pi}{6\eta} \int\limits_0^\infty \!\!
\frac{E(k,0)}{ k^2} {\rm d}k,
\label{41a}
\ende
so that
\beg
\alpha_1=\frac{\pi}{6} {\rm Rm}\ \ell_{\rm corr}^2,
\label{41b}
\ende
which is similar to the result (\ref{41}).
The factor $\pi/6$ also appears in Fig.~\ref{f1}
for small $\rm Pm$ as the value of $I_1$ at the left vertical axis
for $\tau_{\rm corr}=0$.
The same procedure for $\rm Pm=1$, applied to (\ref{38})$_1$, leads to
\beg
\alpha_1=\frac{\pi}{15} {\rm Rm}\ \ell_{\rm corr}^2.
\label{41c}
\ende
Now, one finds the factor $\pi/15$ on the left vertical axis of
Fig.~\ref{f1} (top) for $\rm Pm=1$.
Note that the result  for small $\rm Pm$ exceeds that for $\rm Pm=1$.
For given $\eta$, smaller values of the viscosity $\nu$ lead to {\em
higher values} of the EMF. It is this unexpected behavior that makes
experiments with fluid metals with their small magnetic Prandtl numbers
much-promising.

\begin{figure}[htb]
  \includegraphics[width=8cm]{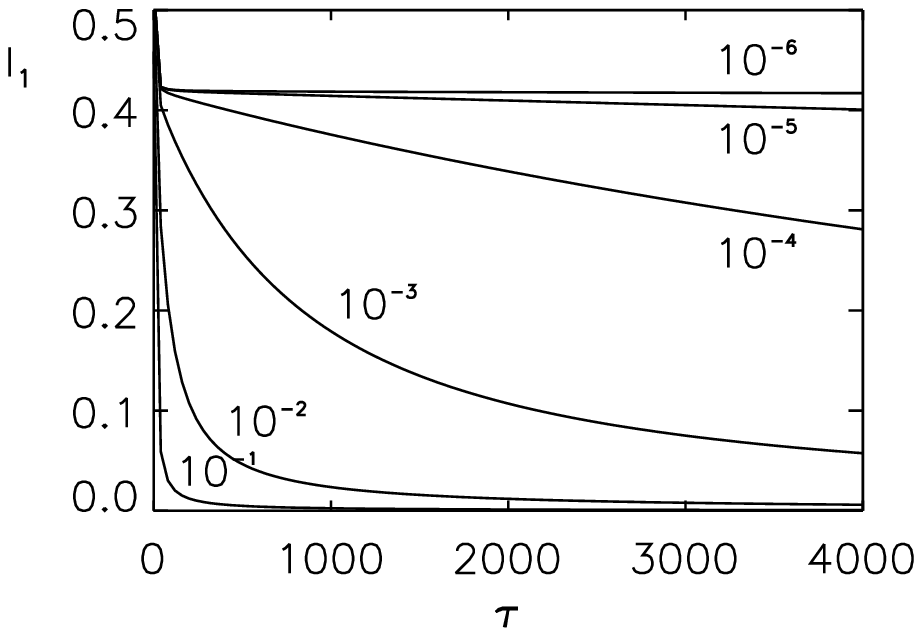}
      \caption{$I_1$ for various $\rm Pm$. Obviously,  for given magnetic diffusivity $\eta$ fluids with much smaller viscosity (like sodium and gallium) are even better qualified for laboratory experiments.}
   \label{f2}
 \end{figure}
While very small values of $\tau_{\rm corr}$ (relative to the magnetic diffusion
time $\tau_{\rm diff}$) represent the high-conductivity limit, a much larger
value of $\tau_{\rm corr}$ represents the low-conductivity limit,
which can be treated by
assuming $w\to 0$ in Eq.~(\ref{18}), which corresponds to
using a $\delta$ function in Eqs.~(\ref{38}).
It directly follows from (\ref{38})$_1$ that
\beg
\alpha_1=\frac{1}{15} \left(1+ \frac{4}{\rm Pm}\right) {\rm Rm}^2\ \ell_{\rm corr}^2,
\label{41c2}
\ende
so that $\rm Pm=1$ leads to
$\alpha_1= \frac{1}{3}{\rm Rm}^2\ \ell_{\rm corr}^2$.
For small magnetic Prandtl number one finds
$
\alpha_1=\frac{4}{15} {\rm Re}\ {\rm Rm}\ \ell_{\rm corr}^2,
$
where $\rm Re= Rm/Pm$ is the Reynolds number. Hence, in the
low-conductivity limit (small $\rm Rm$) the $\alpha$ effect runs with
$\rm Re\cdot  Rm$, while in the high-conductivity limit (large $\rm
Rm$) the  $\alpha$ effect runs with $\rm Rm$. Very similar expressions
also occur if the shear flow is formally replaced by a basic rotation
\cite{Rue78}.

The general expression for finite correlation times,
which is valid between the high and low conductivity limits,
might also be written in the form 
\beg
\alpha_1 = I_1(\tau)\ {\rm Rm}\ \ell_{\rm corr}^2,
\label{41d}
\ende
where $I_1(\tau)$ is given in Fig.~\ref{f1}.
This is the result of a numerical integration using
a spectral function of the form ${\rm exp}(-\tau^2 \omega^2)$
with the dimensionless parameter
\beg
\tau=\eta k^2 \tau_{\rm corr} \simeq 1/{\rm Rm},
\label{tau} 
\ende 
for the aforementioned ratio of the correlation time to the
magnetic diffusion time of the eddies.
$\tau<1$ gives the sector of high conductivity and $\tau>1$ gives the
sector of low conductivity.
One finds that for the given range of $\tau$ for small $\rm Pm$, the
function $I_1$ is nearly uniform (in contrast to the case of $\rm Pm=1$).
The consequence is that, also for lower conductivity, the $\alpha$ effect
only sinks linearly with
$\rm Rm$ rather than quadratically as it is the case for $\rm Pm=1$.
The appearance of the $1/{\rm Pm}$-term in Eq.~(\ref{41c2}) is the
formal reason for this surprising and much-promising behavior.

As a demonstration, Fig.~\ref{f2} shows the behavior of the numerical
integrals for decreasing $\rm Pm$ and very large values of $\tau$.
The integrals are running with  $({\rm Pm}\ \tau)^{-1}$ so that for
${\rm Pm}\ \tau\gg 1$ the relation $\alpha_1\propto {\rm Re\ Rm}$ results.
For small $\rm Pm$, the kinetic Reynolds number is much larger than the
magnetic Reynolds number.
For given magnetic diffusivity, smaller values of the viscosity strongly
enhance  the resulting $\alpha$ effect.

In experiments with liquid metals, $\rm Rm$ is of the order of
0.1...1, so $\tau\simeq 1...10$.
In this regime, and for small $\rm Pm$, the coefficient $I_1$ hardly
changes  with $\tau$ (see Fig.~\ref{f1}).
Its approximate value is 0.4, which is very close to the value $\pi/6$
valid in  the high-conductivity limit.
The reason is that for small $\rm Pm $ the transition of $I_1$ to the
low-conductivity limit only happens at rather high values of $\tau$
(Fig.~\ref{f2}).
In this limit one finds a strong influence of the magnetic Prandtl number.

Similar calculations for  $\alpha_2$ lead to 
\beg
\alpha_2 \simeq -I_2\ {\rm Rm}\ \ell_{\rm corr}^2,
\label{alf2}
\ende
where $I_2$ is also plotted in Fig.~\ref{f1}.
For $\tau$ of the order of unity
and small $\rm Pm$, we find $I_2\lsim I_1\approx0.4$, so
\beg
\alpha_2\simeq - 0.4\ {\rm Rm}\ \ell_{\rm corr}^2. 
\label{42c}
\ende
In the low-conductivity limit ($\tau\gg 1$), we have 
\beg
\alpha_2=-\frac{1}{60} \left(\frac{4}{\rm Pm}-5\right) {\rm Rm}^2\ \ell_{\rm corr}^2, 
\label{42d}
\ende
which   changes its sign at $\rm Pm=0.8$ and yields for   small $\rm Pm$ 
\beg
\alpha_2=-\frac{1}{15} {\rm Rm} \, {\rm Re}\ \ell_{\rm corr}^2 = - \alpha_1. 
\label{42d2}
\ende
Indeed, Fig.~\ref{f1} shows that for large values of $\tau$ and $\rm Pm=1$, $I_2$
is negative (and small), but for $\rm Pm\ll 1$ it is positive ($\simeq 0.4$).

In summary, the plots in Fig.~\ref{f1} reveal an important influence
of the value of the viscosity on the $\alpha$ effect for a given diffusivity.
For small $\rm Pm$, using the low-conductivity limit, the ratio of $I_2$ to
$I_1$ approaches unity, while for $\rm Pm=1$ it is very small.
Note that for $\rm Pm=1$, $\alpha_1$ strongly exceeds $\alpha_2$,
but this is no longer the case for small $\rm Pm$.
The two horizontal components of the $\alpha$ tensor are then of the same
order of magnitude.
The signs of the components are always identical.

It must also be mentioned that the magnetic Prandtl number is much smaller
for liquid metals than what can presently be used in numerical simulations.
Figure~\ref{f1} shows that in numerical simulations, $\alpha_{xx}$
should be smaller than $\alpha_{yy}$, what is not true, however,
for laboratory conditions with their small $\rm Pm$.
In this case it does not matter whether the shear-induced $\alpha$
effect is measured in the streamwise or the spanwise direction.

This finding is in stark contrast to the results of the turbulence model
described in \cite{Ki91} 
for applications to convection zones.
This model works with a very steep frequency spectrum,
$E\propto \delta(\omega)$, and assumes ${\rm Pm}=\tau=1$
for the diffusivities of a postulated small-scale background
turbulence.
This immediately leads to $A_2/A_1=1/20$,
i.e., $\alpha_{xx}/\alpha_{yy}=1/20$.
The considered turbulence model, therefore, yields a strongly dominating $\alpha$
effect in the azimuthal direction (as  also in our approach for $\rm Pm=1$).

It is also obvious that the pumping term does not depend on the shear.
After (\ref{36}), for small $\eta$, it does not run with $1/\eta$.
It is simply
\begin{equation}
\gamma\simeq u^2_{\rm rms} \tau_{\rm corr} \simeq \eta\ {\rm Rm} .
\label{gam}
\end{equation}
It can only be measured if the external field $B_0$ lies in the shear plane.

\section{Numerical simulations}
 
It is straightforward to verify the existence of an $\alpha$ effect in a shear flow using
numerical simulations of non-uniformly forced turbulent shear flows
in Cartesian coordinates.
We perform simulations in a cubic domain of size $L^3$,
so the minimal wave number is $k\equiv k_1=2\pi/L$.
We solve the equations of compressible hydrodynamics
with an isothermal equation of state with constant sound speed $\cs$,
\EQ
{\DD\UU\over\DD t}=SU_x\yyy-\cs^2\nab\log\rho
+\ff+\rho^{-1}\nab\cdot2\rho\nu\SSSS,
\label{dUU}
\EN
\EQ
{\DD\log\rho\over\DD t}=-\nab\cdot\UU,
\label{dlnrho}
\EN
where $\DD/\DD t=\partial/\partial t+\UU\cdot\nab+Sx\;\partial/\partial y$
is the advective derivative with respect to the full velocity field
(including the shear flow), $\UU$ is the departure from the mean shear
flow $(0,Sx,0)$, and $\SSS_{ij}={1\over2}
(\partial_i U_j+\partial_j U_i)-{1\over3}\delta_{ij}\nab\cdot\UU$
is the trace-less rate of strain matrix
(not to be confused with the shear rate $S$).
The flow is driven by a random forcing function $\ff$ consisting of non-helical
waves with wave numbers whose modulus lie in a narrow band around an
average wave number $k_{\rm f}=5k_1$ \cite{HBD04}.
We arrange the amplitude of the forcing function such that
the rms velocity increases with height, while the maximum Mach
number remains below 0.1, so the effects of compressibility are
negligible.
The resulting flow is irregular in space and time and will loosely be
referred to as turbulence.

We use the kinematic test-field method \cite{Sch07} in the Cartesian
implementation \cite{BRRK08} to compute from the simulations
simultaneously the relevant components of the $\alpha$ effect and
turbulent diffusivity tensors, $\alpha_{ij}$ and $\eta_{ij}$.
We do this by solving an additional set of
equations governing the departure of the magnetic field
from a set of given mean fields.
This mean field is referred to as a test field
and is marked by the superscript T.
For each test field $\meanBBT$, we find the corresponding fluctuations
${\bbT={\rm rot}\aaT}$ by solving the inhomogeneous equation
for the corresponding vector potential $\aaT$,
\EQ
{\DDD\aaT\over\DDD t}=-Sa_y^{\rm T}\xxx+
\meanUU\times\bbT+\uu\times\meanBBT+\left(\uu\times\bbT\right)'
+\eta\nabla^2\aaT,
\EN
where $\DDD/\DDD t=\partial/\partial t+Sx\;\partial/\partial y$ is the
advective derivative with respect to the imposed shear flow only
(i.e., without $\UU$), and
${\left(\uu\times\bbT\right)'=\uu\times\bbT-\overline{\uu\times\bbT}}$
is the fluctuating part of $\uu\times\bbT$.
We compute the corresponding mean electromotive force,
${\bm{{\cal E}}^{\rm T}=\overline{\uu\times\bbT}}$, which is then related
to $\meanBBT$ and its curl, $\mu_0\meanJJT={\rm rot}\meanBBT$, via
\EQ
{\cal E}_i^{\rm T}=\alpha_{ij}\meanBT_j-\eta_{ij}\mu_0\meanJT_j.
\label{TFM}
\EN
We use 4 different test fields with $x$ or $y$ components being
proportional to $\sin kz$ or $\cos kz$.
The $x$ and $y$ components of \Eq{TFM} constitute then 8 equations
for the 4 relevant components of $\alpha_{ij}(z,t)$ and $\eta_{ij}(z,t)$.

We adopt periodic boundary conditions in the $y$ direction,
shearing--periodic boundary conditions in the $x$ direction, and
stress-free perfect conductor boundary conditions in the $z$ direction, i.e.,
\EQ
\partial_z u_x=\partial_z u_y=u_z=\aT_x=\aT_y= \partial_z\aT_z=0.
\EN
Numerical resolutions of $64^3$ and $128^3$ mesh points were found
to be sufficient, depending on the value of ${\rm Pm}$.
The {\sc Pencil Code\,} \cite{pencil}
has been used for all calculations.

Simulations are performed for different parameter combinations.
The quantities $S$ and $g_z$ are positive in the calculations presented here,
i.e.\ the basic velocity (\ref{1}) grows in the positive $x$ direction
while the turbulence intensity grows in the positive $z$ direction.
To make contact with laboratory experiments, we focus here on the case
of low conductivity and choose ${\rm Rm}\equiv\urms/\eta\kf=0.2$,
which is consistent with our definition of \Eq{RmDef} with a
Strouhal number of unity, i.e. $\tau_{\rm corr}\urms\kf=1$. 
\begin{figure}[t!]
\includegraphics[width=8cm]{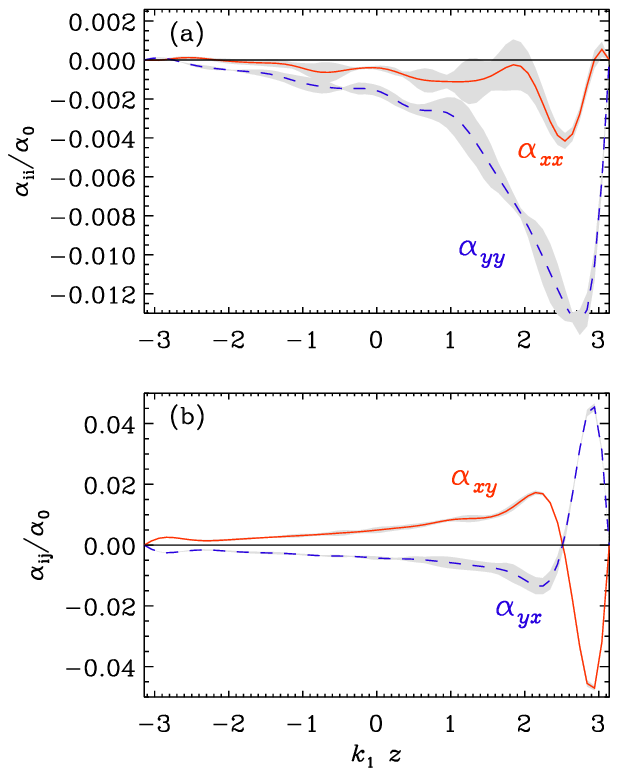}
\caption{Simulations with positive shear ($s=0.2)$:
The numerical values for the diagonal elements of the $\alpha$ tensor (top)
and the off-diagonal elements (bottom).
Error margins are indicated in gray.
 $\rm Pm=1$. $\rm Re=0.2$, $\rm Rm=0.2$.}
\label{f3a}
\end{figure}
\begin{figure}
\includegraphics[width=8cm]{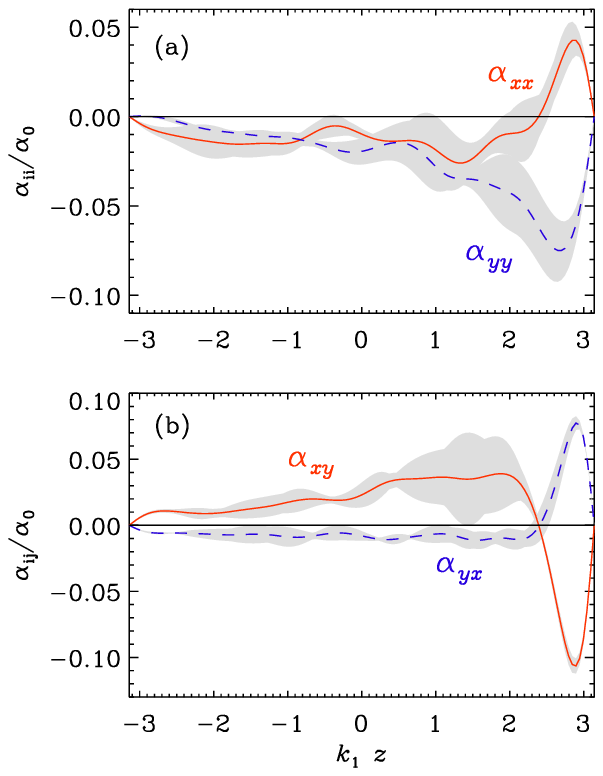}
\caption{The same as in Fig.~\ref{f3a}, but for $\rm Pm=0.1$. $\rm Re=2.5$, $\rm Rm=0.25$.}
\label{f3ab}
\end{figure}
As in earlier work using fully helical turbulence, we present
time averages of
the components of $\alpha_{ij}$ and $\eta_{ij}$ in normalized
form in terms of $\alpha_0=\urms/3$ and $\eta_{\rm T0}=\urms/3\kf$.
Hence, $\alpha_0\, L/\eta_{\rm T0}=L\kf= 10\pi$. 
Error margins are estimated as the largest departure of any one third
of the full times series of  $\alpha_{ij}$ and $\eta_{ij}$.
The shear of the background flow is normalized with the speed of sound, i.e.\
\beg
S= s c_{\rm ac} k_1=  \frac{2\pi s u_{\rm rms}}{{\rm Ma} L},
\label{43a}
\ende
where ${\rm Ma}= u_{\rm rms}/c_{\rm ac}$ is the Mach number.
In the simulations we work with $s=0.2$ and $\rm Ma=0.05$.
One finds
\beg
\frac{\alpha_{yy}}{\alpha_0}\simeq - \frac{\ell_{\rm corr}^2}{L^2}.
\label{43b}
\ende

Following Eqs.~(\ref{35}), both streamwise and spanwise $\alpha$ tensor
components, $\alpha_{yy}$ and $\alpha_{xx}$, should be negative.
When the simulations are done for $\rm Pm=1$, $|\alpha_{yy}|$
should strongly exceed the value of $|\alpha_{xx}|$,
but this is {\em not} expected for $\rm Pm<1$.
Here, the results of two simulations are presented.
The first one for $\rm Pm=1$ with $64^3$ meshpoints has $\rm Rm=0.2$ and $\rm Re\cdot Rm=0.04$, while
the second one  for $\rm Pm=0.1$  with $128^3$ meshpoints has $\rm Rm=0.25$ and $\rm Re\cdot Rm=0.625$.
It is thus possible to find out whether the simulated $\alpha$ effect
runs with $\rm Rm$ (which is almost the same) or with $\rm Rm\cdot Re$ (which   differs by a factor
of 10) in both simulations.
In both cases $k_{\rm f}/k_1=5$ so that 10 cells can  exist in the vertical
direction and, therefore,  $ \ell_{\rm corr}^2/{L^2} \simeq  0.01$.

As predicted, Fig.~\ref{f3a}(a) for $\rm Pm=1$ shows $\alpha_{yy}$
to be dominant and 
both diagonal components of $\alpha$ as basically  negative. The 
 amplitude of ${\alpha_{yy}}/\alpha_0$ is about  0.01 in accordance to  Eq.~(\ref{43b}) which also leads to  $|\alpha_{yy}|/\alpha_0 \simeq  0.01$.

For $\rm Pm=1$, $I_2$ is strongly reduced relative to $I_1$ so that the
small amplitude of $\alpha_{xx}$ in Fig.~\ref{f3a}(a) becomes understandable.
For smaller magnetic Prandtl number, this reduction does not exist and both
$\alpha$ components are of similar amplitude. Close to the upper endplate the intensity stratification changes its sign (due to the boundary conditions) and also a change of the sign of the $\alpha$ effect can be observed there (see Fig.~\ref{f3ab})(a).
Without this exception the simulations also confirm that the  signs  only depend on the sign
of the product $g_z S$, as formulated in the relations (\ref{35}).

Moreover, again as predicted, the amplitudes of the diagonal elements of
the $\alpha$ tensor increase for decreasing magnetic Prandtl number.
In the middle of the channel, the amplitudes of the $\alpha$ components
differ by a factor of 10 which exceeds the ratio 1.25 of the two $\rm Rm$
by almost an order of magnitude.

Next, the off-diagonal components of $\alpha_{ij}$ are considered;
see Figs.~\ref{f3a}(b) and \ref{f3ab}(b).
As expected, we have $\alpha_{yx}\approx-\alpha_{xy}$, which corresponds to
a turbulent pumping velocity in the $z$ direction.
This velocity is negative for $k_1 z<2.5$, corresponding to downward transport,
i.e., down the gradient of the turbulent intensity, as expected \cite{KR80}.
Near the top of the domain, the gradient of the turbulent intensity is
reversed and so is the sign of the pumping velocity $\alpha_{yx}$.
The simulations with $\rm Pm=0.1$ and $1$ lead to the result
$\alpha_{yx}/\alpha_0\simeq - {\rm Rm}/10$ (Figs.~\ref{f3a}b and \ref{f3ab}b).

\begin{figure}
\includegraphics[width=8cm]{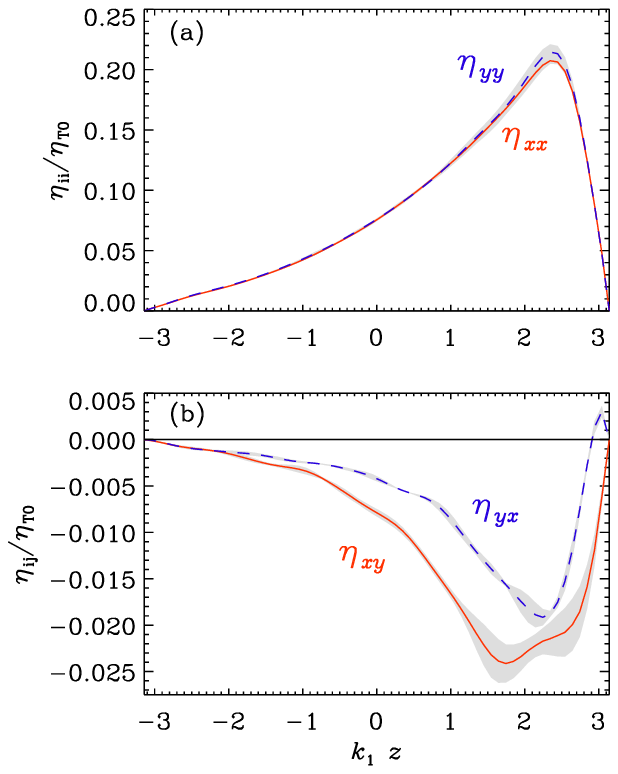}
\caption{Simulations with $s=0.2$ for the shear-induced elements of the
$\eta$ tensor.
Top: the horizontal eddy diffusivities;
bottom: the two shear--current terms $\eta_{xy}$ and $\eta_{yx}$.
Error margins are indicated in gray.
$\rm Pm=1$. $\rm Re=0.2$, $\rm Rm=0.2$.}
\label{f3b}
\end{figure}

\begin{figure}
\includegraphics[width=8cm]{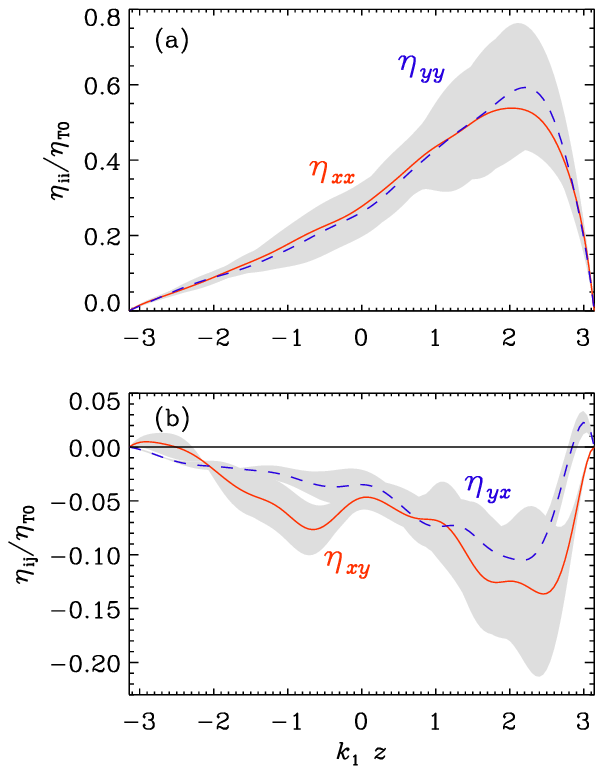}
\caption{The same as in Fig. \ref{f3b}, but for $\rm Pm=0.1$.  $\rm Re=2.5$, $\rm Rm=0.25$.}
\label{f3bb}
\end{figure}

\section{The diffusivity tensor}
The numerical test-field procedure also allows the simultaneous
calculation of  the components of the $\eta$ tensor for the same
simulation with its vertical stratification of the turbulence intensity.
This knowledge is important for the discussion of the question whether
turbulence in shear flows can be used for dynamo self-excitation of
large-scale magnetic fields.
Despite the completely different roles played by the spanwise and
streamwise directions, we find the two relevant diagonal components of the diffusivity tensor 
to be nearly equal, i.e., $\eta_{xx}\approx\eta_{yy}$
(Figs.~\ref{f3b}a and \ref{f3bb}a).
This strikingly high degree of isotropy of the turbulent diffusivity
in the $xy$ plane has already been noticed in earlier simulations of
the diffusivity in unstratified turbulent shear flows \cite{B05,RK06,BRRK08}.
The maximum of $\etaT=(\eta_{xx}+\eta_{yy})/2$ is about 20\% of the reference
value $\eta_{0}$, which agrees with the fact that for $\Rm\ll1$,
$\etaT/\eta_{0}\simeq\Rm$ \cite{Sur}.
For larger $\rm Rm$ (or for smaller $\rm Pm$) one finds slightly larger
numerical values for the eddy diffusivity (Fig.~\ref{f3bb}a).

The data in Figs.~\ref{f3a}(a) to \ref{f3bb}(a) for $\rm Pm=0.1$ and $1$
lead to a value of about unity for the normalized $\alpha$ effect,
$C_\alpha= \alpha\,L/\eta_{\rm T}$, which is also typical for rapidly
rotating convection \cite{KKB09}.
A comparison with the slab-dynamo calculation in \cite{RK06} leads to
$C_\alpha\simeq 10$ as required for self-excitation of the magnetic fields.
This condition is not fulfilled for the present simulations.

For the off-diagonal components of the $\eta$ tensor 
for non-stratified shear flows one finds 
\beg
\eta_{xy}= \eta_x S,\quad \eta_{yx}= \eta_y S,
\label{sc3}
\ende
i.e., both are linear in $S$.
The calculation of a simple slab dynamo model shows self-excitation for
sufficiently large positive $\eta_y$.
From quasilinear theory we know, however, that $\eta_y$ is
negative-definite \cite{RK06}.
For positive shear, the coefficient $\eta_{yx}$ is therefore expected
to be negative.
This result has also been confirmed for $\Rm\leq 200$
for unstratified turbulence \cite{BRRK08}.

The same sign and the same linear dependence of $\eta_{yx}$ on $S$ also
holds for $\eta_{xy}$, but only for $\rm Pm$ of order unity and
in the low conductivity limit.
Both conditions are fulfilled for the present simulations.
Experiments for liquid metals, however, concern much smaller magnetic Prandtl
numbers, for which $\eta_{xy}$ is expected to be positive.

Our numerical simulations for stratified turbulence and with positive shear
and $\rm Pm\leq 1$ also produce negative values for both $\eta_{xy}$ and
$\eta_{yx}$ (Figs.~\ref{f3b}b and \ref{f3bb}b).
The possibility of dynamo action in such non-helical shear flows
\cite{RK03,BRRK08,Yousef} can therefore not be explained by the so-called 
shear--current effect.

It is also shown that the vertical stratification of the turbulence
intensity does not basically modify the known findings about the eddy
diffusivity tensor.
Only experiments can finally provide the sign of $\eta_{x}$ as simulations
for such small $\rm Pm$ are not usually possible.

\section{Shear flow electrodynamics}
Following relations (\ref{35}), the electromotive force across the channel is
\beg
{\cal E}_x= \alpha_2 g_z S B_0,
\label{44}
\ende
so that the potential difference $\delta\varPhi$ between the walls
with distance $D$ is $\delta\varPhi= \alpha_2 g_z S D B_0$ so that
\beg
\delta\varPhi \simeq - 0.5\,{\rm Rm}\,
\ell_{\rm corr}^2 g_z U B_0
\simeq  - {\rm Rm}\ \lambda^2 \ L U B_0 
\label{45}
\ende
with $I_2\simeq 0.5$ for small $\rm Pm$ (Fig. \ref{f1}, bottom), with $g_z=2/L$ as the vertical scale height of the turbulence stratification and with the ratio $\lambda=\ell_{\rm corr}/L$. The amplitude of the mean shear flow is $U$. Note that surprisingly the width $D$ of the channel does not appear in   (\ref{45}) and even  the height $L$  has only a weak influence.
Hence,
\beg
\delta\varPhi\simeq 10\ {\rm Rm}\, \lambda^2 \left[\frac{L}{10\,{\rm cm}}\right] \left[\frac{U}{{\rm m/s}}\right] \left[\frac{B_0}{\rm kgauss}\right]
\label{46}
\ende
(in mVolt) so that with (say) $\lambda\simeq 0.1$ and a channel height
of 50\,cm, a shear flow of 1 m/s subject to a magnetic field of 1\,kgauss
would lead to a potential difference of
\beg
\delta\varPhi\simeq 0.5 \ {\rm Rm}\ [\rm mV].
\label{47}
\ende
For the (maximal) value of $\rm Rm\simeq 1$ ($u_{\rm rms} \simeq 1$m/s
and $\ell_{\rm corr}\simeq 5$ cm) the channel should thus provide
a potential difference of 0.5\,mV between the side walls by the action of the $\alpha$ effect along a spanwise magnetic field. 
These numbers are quite similar to those of the Riga experiment
\cite{St67,KR80}.
The basic difference is that in our shear flow the helicity
is not prescribed but it is self-consistently produced by the
interaction of the stratified turbulence with the background shear.

\smallskip
\section{Conclusions}
Laboratory studies of homogeneous dynamos are still in their infancy.
The only working dynamo where the flow pattern is not strongly constrained
by pipes or container walls is the experiment in Cadarache \cite{Cada},
where, however, the effects of soft iron play an important and not
well understood role \cite{gies12}. 
The present proposal of measuring the $\alpha$ effect in an {\em unconstrained}
turbulent flow would therefore be a major step forward.
In such an experiment, the pseudoscalar necessary for producing helicity
comes from the stratification of turbulent intensity giving rise to
a polar vector and the vorticity associated with the shear flow
giving rise to an axial vector.
Thus, the basic effects in the theory of turbulent dynamos,
which are usually considered as special properties of {\em rotating} and
stratified fluids, can also be found for the plane-shear flows,
i.e.\ without global rotation.

The present work yields  a detailed prediction about the sign and
magnitude of the components of both $\alpha$ and $\eta$ tensors.
It may motivate the construction of a suitable experiment
using liquid metals to achieve a measurable $\alpha$ effect.
The necessary vertical stratification of turbulence intensity
must be experimentally imitated using grids with nonuniform mesh sizes
and/or walls of increasing/decreasing roughness in the vertical direction.

We have shown that in stratified turbulence driven in a plane shear flow,
a measurable $\alpha$ effect should exist.
Here, the key problem is the smallness of the magnetic Prandtl number.
For $\rm Pm\leq 1$, the quasilinear theory and the possible nonlinear
numerical simulations lead to very similar results.
With the quasilinear theory we have shown that,
even for fluids with very small magnetic Prandtl numbers,
stratified shear flow turbulence leads to an $\alpha$ effect
that can be realized in an experiment with liquid metals such as sodium
($\rm Pm\simeq 10^{-5}$) or gallium ($\rm Pm\simeq 10^{-6}$).
Such small magnetic Prandtl numbers cannot be simulated with
present-day numerical codes.

In fact, it may not be possible that such flows could produce a
supercritical dynamo in the conceivable future.
Nevertheless, even in the subcritical case, an $\alpha$ effect should be
measurable, which would thus open the possibility of detailed comparisons
between theory, simulations and experiments.
Once such a comparison is possible, there will be more details that
should be investigated.
One of them concerns the modifications of the results in the presence
of imperfect scale separation in space and time.
For oscillatory dynamos, this effect can significantly lower the
excitation conditions for the dynamo compared to standard mean-field
estimates \cite{RB12}.

\acknowledgments
Financial support from the European Research Council under the AstroDyn
Research Project 227952, the Swedish Research Council under the grants
621-2011-5076 and 2012-5797, as well as the Research Council of Norway
under the FRINATEK grant 231444 are gratefully acknowledged.


\newcommand{\ygafd}[3]{, Geophys.~Astrophys.~Fluid~Dynam. {\bf #2}, #3 (#1).}
\newcommand{\yapj}[3]{, Astrophys.~J. {\bf #2}, #3 (#1).}
\newcommand{\ypre}[3]{, Phys.~Rev.~E {\bf #2}, #3 (#1).}
\newcommand{\yprl}[3]{, Phys.~Rev.~Lett. {\bf #2}, #3 (#1).}
\newcommand{\ymn}[3]{, Mon.~Not.~R.~Astron.~Soc. {\bf #2}, #3 (#1).}
\newcommand{\yan}[3]{, Astron.~Nachr. {\bf #2}, #3 (#1).}



\begin{thebibliography}{}
\bibitem{KR80}
F. Krause and K.-H. R\"adler, {\it Mean-Field Magnetohydrodynamics and Dynamo Theory} (Pergamon Press, Oxford, 1980).

\bibitem{B05}
A. Brandenburg\yan{2005}{326}{787}

\bibitem{RH04}
G. R\"udiger and R. Hollerbach, {\it The Magnetic Universe: Geophysical and Astrophysical Dynamo Theory} (Wiley-VCH, Berlin, 2004).


\bibitem{St12}
V. Noskov, S. Denisov, R. Stepanov, and P. Frick, Phys. Rev. E, {\bf 85}, 016303 (2012).

\bibitem{St67}
M. Steenbeck, I.~M. Kirko, A. Gailitis, A.~P. Klawina, F. Krause, I.~J. Laumanis, and O.~A. Lielausis, Monats. Dt. Akad. Wiss. {\bf 9}, 714 (1967).

\bibitem{St06}
R. Stepanov, R. Volk, S. Denisov, P. Frick, V. Noskov, and J.-F. Pinton, Phys. Rev. E {\bf 73}, 046310  (2006).

\bibitem{F08}
P. Frick, S. Denisov, V. Noskov, and R. Stepanov, Astron. Nachr. {\bf 329}, 706 (2009).

\bibitem{RK06}
G. R\"udiger and L.~L. Kitchatinov, Astron. Nachr. {\bf 327}, 298 (2006).

\bibitem{Sch07}
M. Schrinner, K.-H. R\"adler, D. Schmitt, M. Rheinhardt, and U. R. Christensen\ygafd{2007}{101}{81}

\bibitem{RS06}
K.-H. R\"adler and R. Stepanov\ypre{2006}{73}{056311}

\bibitem{Rue78}
G. R\"udiger\yan{1978}{299}{217}

\bibitem{Ki91}
L. L. Kichatinov, Astron. Astrophys. {\bf 243}, 483 (1991).

\bibitem{HBD04}
N. E. L. Haugen, A. Brandenburg, and W. Dobler\ypre{2004}{70}{016308}

\bibitem{BRRK08}
A. Brandenburg, K.-H. R\"adler, M. Rheinhardt, and P.J. K\"apyl\"a\yapj{2008}{676}{740}

\bibitem{pencil}
http://pencil-code.googlecode.com

\bibitem{Sur}
S. Sur, A. Brandenburg, and K. Subramanian\ymn{2008}{385}{L15}

\bibitem{KKB09}
P. K\"apyl\"a, M. Korpi,  and A. Brandenburg, Astron. Astrophys. {\bf 500}, 633 (2009).

\bibitem{Yousef}
T. A. Yousef, T. Heinemann, A. A. Schekochihin, N. Kleeorin, I. Rogachevskii, A.B. Iskakov, S. C. Cowley, and J. C. McWilliams\yprl {2008}{100}{184501}

\bibitem{RK03}
I. Rogachevskii and N. Kleeorin\ypre{2003}{68}{036301}

\bibitem{VB97}
E. T. Vishniac and A. Brandenburg\yapj{1997}{475}{263}

\bibitem{Cada}
R. Monchaux, M. Berhanu, M. Bourgoin, M. Moulin, Ph. Odier, J. -F. Pinton, et al.\yprl{2007}{98}{044502}

\bibitem{gies12}
A. Giesecke, C.  Nore, F.  Stefani, G. Gerbeth, et al., NJP {\bf 14}, 3005 (2012).

\bibitem{RB12}
M. Rheinhardt and A. Brandenburg\yan{2012}{333}{80}

\end{thebibliography}
\end{document}